\documentclass[12pt]{article}
\usepackage{pdproc,epsfig} 

 \newcommand{\mn}{\Delta m_{21}^2}
\newcommand{\mt}{\Delta m_{31}^2}
\newcommand{\si}{s_{12}}
\newcommand{\sn}{s_{23}}
\newcommand{\st}{s_{13}}
\newcommand{\ci}{c_{12}}
\newcommand{\cn}{c_{23}}

\begin{document}

\renewcommand{\thefootnote}{\alph{footnote}}
  
\title{
NEUTRINO'S NON-STANDARD INTERACTIONS; ANOTHER EEL UNDER A WILLOW?\protect\footnote{
Written version of a talk presented at 
XIII International Workshop on Neutrino Telescopes, Istituto Veneto di Scienze, Lettere ed Arti, Venice, Italy, March 10-13, 2009. } }

\author{HISAKAZU MINAKATA}

\address{Department of Physics, Tokyo Metropolitan University, \\
1-1 Minami-Osawa, Hachioji, Tokyo 192-0397, Japan\\
 {\rm E-mail: minakata@phys.metro-u.ac.jp}}




\abstract{
I report some progress that occurred since NO-VE 08 in the field of non-standard interactions (NSI) of neutrinos. After briefly reviewing theoretical developments, I give a summary of the two works in which I was involved. Firstly, we have formulated a perturbative framework to illuminate the global features of neutrino oscillations with NSI, aiming at exploring method for determination of the standard mixing and the NSI parameters. We have recognized that the parameter degeneracy prevails with an extended form which involves the NSI elements. Furthermore, a completely new type of degeneracy is shown to exist. The nature of the former degeneracy is analyzed in detail in the second work. The work is primarily devoted to analyze the problem of discriminating the two CP violation, one due to the lepton Kobayashi-Maskawa phase and the other by phase $\phi$ of the NSI elements. We have shown that the near (3000 km)$-$far (7000 km) two detector setting in neutrino factory does have the discrimination capability and is sensitivities to CP violation due to NSI to $\vert \varepsilon_{e\mu} \vert$ to $\simeq$ several $\times 10^{-4}$ in most of the region of $\delta$ and $\phi_{e \mu}$. 
}
   
\normalsize\baselineskip=15pt

\section{Introduction}

The question I would like to address in my talk is: 
Are there something terribly new in neutrino properties after the discovery 
of neutrino masses and lepton mixing\cite{MNS}? 
The Japanese saying in the subtitle is meant to be that.\footnote{
Though the original saying in Japan is ``Another loach under a willow?'' 
I decided to keep the modified version because eel is much more familiar 
to us, and in particular thanks to a great support to the eel version by the 
spokesman of IceCube collaboration. 
}
%
Clearly this is an extremely interesting question. 
But, I must start with a cautionary remark. 

What is the natural time scale for discovery of something extremely new 
in neutrino properties? 
Let us look back the history to obtain a hint for answering the question. 
It took more than 60 years from the Meitner-Hahn measurement of 
electron energy spectrum in nuclear beta decay in 1911 to the discovery of 
NC reaction in 1973 \cite{Haidt}. 
From the discovery of neutrino itself in 1953 by Reines and Cowan 
\cite{Reines} to the discovery of neutrino mass and lepton flavor mixing 
by Super-Kamiokande in 1998 \cite{SKatm-evidence} 
needed 45 years. 
Thus, the right time scale, as history tells us, is $\sim$ 50 years. 
It implies a warning; 
What people think about the possible candidates for 
``terribly new in the neutrino properties'' at the right time can be very 
different from those we consider today. 
Though I have to speak within the scope that I can think of today 
but this point has to be kept in mind as a word of caution.

It also has to be remarked that the scope of my presentation is very limited; 
Though I restrict myself into the so called non-standard interactions (NSI) 
of neutrinos \cite{wolfenstein,valle,guzzo,grossmann} in my talk, 
``new in the neutrino properties'' may include 
various often more radical possibilities such as:  
departure from three flavor mixing, sterile neutrinos, violation of 
fundamental symmetries like CPT. 
See e.g., \cite{concha-maltoni} for a status summary for these more 
exotic possibilities.

\section{Non-Standard Interactions of Neutrinos}

It has been proposed that neutrinos might possess yet unknown 
new neutrino interactions \cite{wolfenstein,valle,guzzo,grossmann}. 
Today we have Standard Model of particle physics, one of the most 
successful theories in physics. Therefore, when we discuss NSI it is 
natural (and to a large extent mandatory) to talk about it in a language 
of higher dimensional operators \cite{weinberg}.

Suppose that there exist a new physics at energy scale $M_{NP}$, 
which I assume to be greater than $\sim 1$ TeV, but not too much larger 
than this value. 
I assume the type of higher-dimentional operators for effective interactions 
of neutrinos with matter \cite{grossmann,berezhiani} 
\begin{eqnarray}
{\cal L}_{\mbox{eff}}^{\mbox{NSI}} = 
-2\sqrt{2}\, \varepsilon_{\alpha\beta}^{fP} G_F
(\overline{\nu}_\alpha \gamma_\mu P_L \nu_\beta)\,
(\overline{f} \gamma^\mu P f),
\label{NSI-dim6}
\end{eqnarray}
where $G_F$ is the Fermi constant, and 
$f$ stands for the index running over fermion species in the earth, 
$f = e, u, d$, in which we follow the conventional notation.
$P$ stands for a projection operator and is either
$P_L\equiv \frac{1}{2} (1-\gamma_5)$ or 
$P_R\equiv \frac{1}{2} (1+\gamma_5)$. 
Given the dimension six operator in (\ref{NSI-dim6}) and 
because we normalize the operator with Fermi constant  $G_F$, 
$\varepsilon_{\alpha \beta}$ must be of the order of 
$(M_{W} / M_{NP} )^2 \sim$ 0.01 ($10^{-4}$) if $M_{NP} = 1 (10)$ TeV. 
If we have to go to dimension eight operators their effective strength 
would be at most $(M_{W} / M_{NP} )^4 \sim 10^{-4}$ even for 
$M_{NP} = 1$ TeV. 
The off-diagonal elements may have further suppression.

Since I gave a talk on NSI last year in Venice \cite{NO-VE08-mina}, 
I will restrict myself into developments that occurred after 
NO-VE 08 to show that the field is moving. 
The rest of my report has three parts: 
In section 3 I review the recent development in the theory of NSI. 
From section 4 I change gear to NSI effect in propagation in matter. 
In section 5 I discuss perturbative treatment of the system with NSI.  
Sections 6 and 7 are devoted to further clarifying the properties of the 
system and to discuss the question of discriminating two kind of 
CP violation, one from the lepton Kobayashi-Maskawa 
phase \cite{KM} and the other from phases of the NSI elements, 
the problem discussed in \cite{concha1}. 
My presentation in the last three sections will be based on 
the two recent papers \cite{NSI-perturbation,NSI-2phase}. 
The latter work is a natural continuation of our previous work \cite{NSI-nufact}.

\section{Recent Development in the Theory of NSI}

Let me start by reviewing the development in the theory of NSI that 
occurred very recently. Since long time ago, it has been noticed 
\cite{berezhiani} that phenomenological study 
with NSI of the type (\ref{NSI-dim6}) has a potential caveat. 
To get to the point, let us agree on the following understanding: 
At a high-energy scale where NSI originates 
the $SU(2) \times U(1)$ gauge invariance holds. Then, 
the left-handed neutrino field in the operator (\ref{NSI-dim6}) must be 
elevated into the lepton doublet of $SU(2)$. 
When we require this an obvious problem occurs; The resultant four charged 
lepton operators have to obtain severe constraints from experiments. 
The most stringent is the one imposed by the branching ratio of 
$\mu \rightarrow eee$, $BR(\mu \rightarrow eee) \leq 10^{-12}$ 
\cite{kuno-okada}, which would yield the constraint 
$\vert \varepsilon_{e \mu} \vert \leq 10^{-6}$. 
Of course, nothing is wrong with it. 
But, we would like to avoid this because we are interested in 
observable effects in near (or even remote) future neutrino experiments. 
Therefore, people looked for the possible higher dimensional 
operators which are free from the charged lepton constraints.

Some candidates which were discussed by people 
(see, for example, \cite{bilenky-santamaria,bergmann-etal,enrique1}) 
are: 
\begin{equation}
\mathcal O^a_6 = 
(\bar L_\gamma i \tau_2 L^c_\alpha)(\overline{L^c_\beta} i \tau_2 L_\delta) 
\label{dim6}
\end{equation}
for dimension six operator where $L^c = C \bar L^T$ and $C$ is the 
charge conjugation operator. 
For dimension eight operators they are of the type 
\begin{equation}
\mathcal O^a_8 = 
(\bar{L}_{\beta}\gamma_{\rho} L_{\alpha})(\bar{L}_{\delta} \tilde{H}) \gamma^{\rho}
(\tilde{H}^\dagger L_{\gamma}). 
\label{dim8}
\end{equation}
See, for example, \cite{berezhiani,davidson,enrique1,gavela-etal,enrique2}  
for relevant references. 
Intuitive understanding of (\ref{dim8}) is that the Higgs field v.e.v. projects 
out only the neutrino component of left-handed doublet. 
The meaning of (\ref{dim6}) becomes clear by writing it in 
a form with obvious antisymmetry in flavor space \cite{enrique2}, 
\begin{equation}
2\mathcal O^a_6 = 
(\bar{\ell}_{\alpha} \gamma^\mu \ell_{\beta})(\bar{\nu}_\gamma \gamma_\mu \nu_{\delta})+
(\bar{\ell}_{\gamma} \gamma^\mu \ell_{\delta})(\bar{\nu}_\alpha \gamma_\mu \nu_{\beta})-
(\bar{\ell}_{\alpha} \gamma^\mu \ell_{\delta})(\bar{\nu}_\gamma \gamma_\mu \nu_{\beta})-
(\bar{\ell}_{\gamma} \gamma^\mu \ell_{\beta})(\bar{\nu}_\alpha \gamma_\mu \nu_{\delta}) 
\end{equation}
which implies 
\begin{equation}
\label{antisymm}
\varepsilon^{\alpha\beta}_{\gamma\delta} = 
-\varepsilon^{\gamma\beta}_{\alpha\delta} = 
-\varepsilon^{\alpha\delta}_{\gamma\beta} =
\varepsilon^{\gamma\delta}_{\alpha\beta}. 
\end{equation}
The antisymmetric nature prohibits, for example, 
$\varepsilon^{e e}_{e \mu}$ which would produce NSI effects in neutrino propagation in matter.

Recently, this problem of searching for higher dimensional operators 
without charged lepton constraints has come to conclusion; 
It has been proved that the above two possibilities are unique in 
dimension six and eight operators, respectively, if one wants to avoid 
the charged lepton constraints at the tree level \cite{gavela-etal}.

Now, the question is:
Is avoiding the tree level constraint sufficient to be free from the 
charged lepton constraints?
It is known that the answer is NO \cite{davidson}.
Namely, the dressing by $W$ and $Z$ bosons can produce four 
charged lepton processes which leads to highly restrictive bound in 
some channels, in particular on $\vert \varepsilon^{e e}_{e \mu} \vert$. 
See  \cite{davidson} (new version) for details of the type of bound, and 
for a summary of the other constraints on NSI.

It turned out, however, that it was {\em NOT} the end of the story. 
Biggio, Blennow and Fernandez-Martinez \cite{enrique2} have recently 
pointed out that the bounds on NSI have to be relaxed to a large extent, 
a factor of $\sim 10^{4}$. 
Because of the antisymmetric nature of the dimension six operator (\ref{dim6}), 
the contributions of diagrams with different flavor indices tend to 
cancel and add up to zero in the limit of neglecting the lepton masses.\footnote{
Here, I restrict my discussion into the dimension six operators. 
But, it is fair to note that this discussion is much more relevant for the 
dimension eight operators. See \cite{enrique2}. 
}
%
Turning on lepton masses leaves the contribution of the order of 
$\left( m_{\ell} / M_{W} \right)^2$. 
Notice that it is the unique dimension six operator which is free from the 
tree-level four charged lepton counter part so that we have to live with it. 
Another significant feature is that the bound on $\varepsilon^{e e}_{e \mu}$ 
goes away, or in other word, it must vanish by construction of the 
operator (\ref{dim6}).  

I would like to note here that an important feature is hidden behind 
my shallow description of their results; 
SU(2) gauge invariance. 
That is, imposing the gauge invariance is essential to obtain gauge 
invariant results of logarithmically divergent terms of the one-loop diagrams. 
Go to the original reference \cite{enrique2} for more complete understanding, 
in particular on the meaning of quadratically divergent terms.

\section{NSI in Neutrino Propagation in Matter}

In the rest of my presentation I discuss NSI effects in neutrino propagation 
in matter. It should be remarked, however, that NSI effects are present 
also in production and detection processes of neutrinos, so that my 
discussion is obviously incomplete.

To summarize its effects on neutrino propagation it is customary 
to introduce the $\varepsilon$ parameters, which are defined as
$\varepsilon_{\alpha\beta} \equiv \sum_{f,P} \frac{n_f}{n_e}
\varepsilon_{\alpha\beta}^{fP}$, 
where $n_f$ is the number density of the fermion species $f$ in matter. 
Notice that only the vector combination of the NSI can be probed when we 
discuss neutrino propagation in matter. 
Approximately, the relation 
$\varepsilon_{\alpha\beta} \simeq \sum_{P}
\left(
\varepsilon_{\alpha\beta}^{eP}
+ 3 \, \varepsilon_{\alpha\beta}^{uP}
+ 3 \, \varepsilon_{\alpha\beta}^{dP}
\right)$  
holds because of a factor of $\simeq$3 larger number of 
$u$ and $d$ quarks than electrons in iso-singlet matter. 
Notice that with the dimension six operator (\ref{dim6}) part of 
the first term, the ones from $\varepsilon^{e e}_{e \mu}$ and 
$\varepsilon^{e e}_{e \tau}$ is absent. 
I, however, choose to proceed with a generic framework. 

Using the $\varepsilon$ parameters the neutrino evolution equation 
which governs the neutrino propagation in matter is given as  
\begin{eqnarray} 
i {d\over dt} \left( \begin{array}{c} 
                   \nu_e \\ \nu_\mu \\ \nu_\tau 
                   \end{array}  \right)
 = \frac{1}{2E} \left[ U \left( \begin{array}{ccc}
                   0   & 0          & 0   \\
                   0   & \Delta m^2_{21}  & 0  \\
                   0   & 0           &  \Delta m^2_{31}  
                   \end{array} \right) U^{\dagger} +  
                  a \left( \begin{array}{ccc}
            1 + \varepsilon_{ee}     & \varepsilon_{e\mu} & \varepsilon_{e\tau} \\
            \varepsilon_{e \mu }^*  & \varepsilon_{\mu\mu}  & \varepsilon_{\mu\tau} \\
            \varepsilon_{e \tau}^* & \varepsilon_{\mu \tau }^* & \varepsilon_{\tau\tau} 
                   \end{array} 
                   \right) \right] ~
\left( \begin{array}{c} 
                   \nu_e \\ \nu_\mu \\ \nu_\tau 
                   \end{array}  \right)
\label{general-evolution}
\end{eqnarray}
where $U$ is the MNS matrix \cite{MNS}, and
$a\equiv 2 \sqrt 2 G_F n_e E$ \cite{wolfenstein} where $E$ is the
neutrino energy and $n_e$ denotes the electron number density along
the neutrino trajectory in the earth.  $\Delta m^2_{i j} \equiv
m^2_{i} - m^2_{j}$ with neutrino mass $m_{i}$ ($i=1-3$).
Notice that the phase of $\varepsilon$ parameters may provide new 
source of CP violation \cite{concha1}. 
Another important point is that complexity of the system in (\ref{general-evolution}) 
would lead to confusion in determination of the mixing and the NSI parameters 
\cite{confusion1,confusion2}.

\section{Perturbation Theory of Neutrino Oscillation with NSI}

Obviously, I am a newcomer to the field of NSI. 
When I started to work on this topics I tried to understand the features 
of neutrino oscillations with NSI. 
Alas, I found that not so many things are known. 
The questions I would like to know the answer were:

\begin{itemize}

\item
From the experience in neutrino oscillation with standard interaction (SI) 
I would expect that the appearance channel 
$\nu_{e} \rightarrow \nu_{\mu}$ (or, $\nu_{e} \rightarrow \nu_{\tau}$) 
has great sensitivities to tiny effects of NSI. 
Then, the natural question is; 
Which NSI elements of $\varepsilon_{\alpha \beta}$ in (\ref{general-evolution}) 
give the dominant contribution to $P(\nu_{e} \rightarrow \nu_{\mu})$?
Or, more concretely, how large is the contribution of e.g., 
$\varepsilon_{e \mu}$, $\varepsilon_{e \tau}$, and 
$\varepsilon_{\mu\tau}$ to $P(\nu_{e} \rightarrow \nu_{\mu})$?

\item
What about the disappearance channels though they may be less attractive? 
Namely, what is the size of contributions of $\varepsilon_{e e}$ in 
$P(\nu_{e} \rightarrow \nu_{e})$?
What is the relative importance of 
$\varepsilon_{e \mu}$, $\varepsilon_{e \tau}$, 
$\varepsilon_{\mu \mu}$, and $\varepsilon_{\mu \tau}$ in 
$P(\nu_{\mu} \rightarrow \nu_{\mu})$?

\end{itemize}

I was amazed by the fact that apparently nobody knew the answers 
to these questions!\footnote{
It might be a too strong statement, given the fact that so many people 
are working in this field. 
Any comments are welcome. 
In fact, it appears that answer to these questions were known at least partly 
by Jacobo Lopez-Pavon in UAM, Madrid, though the result was unpublished. 
}
%
I believe that the questions are not only due to academic interests. 
It is because I think that treating the full system (\ref{general-evolution}) 
is really necessary. 
Though people (including myself) do make approximations of ignoring 
some elements keeping only a few of them, but they do so without good reasons. 
It is even more so now because most of the stringent bounds on 
$\varepsilon_{\alpha \beta}$ based on lepton processes, 
the model-independent ones, are gone.
Only when we recognize the correct theory at high energy scale we 
can be sure that the approximation he/she is making is the correct one.

\subsection{$\epsilon$ Perturbation theory }

To answer these questions and to have a global bird-eye view of 
neutrino oscillation with NSI we have formulated a perturbative 
framework \cite{NSI-perturbation}. 
Unfortunately, there is no unique framework because 
we still do not know the value of $\theta_{13}$, 
though the bound on it exists \cite{CHOOZ}. 
The only parameter which we know to be small is the ratio 
$\frac{ \Delta m^2_{21} } { \Delta m^2_{31} } \simeq 0.03$. 
Therefore, we take an ansatz 
\begin{eqnarray} 
\epsilon \equiv 
\frac{ \Delta m^2_{21} } { \Delta m^2_{31} } 
\sim s_{13} 
\sim \varepsilon_{\alpha \beta} 
\sim 10^{-2}
\hspace{0.5cm}
(\alpha, \beta = e, \mu, \tau) 
\label{def-epsilon}
\end{eqnarray}
to formulate our perturbation theory, which we called 
the ``$\epsilon$ perturbation theory'' in \cite{NSI-perturbation}.
In doing so I assume 
$\frac{ a } { \Delta m^2_{31} }$ is of order unity, anticipating very long-baseline neutrino experiments such as neutrino factory \cite{nufact}, or the beta beam 
\cite{beta}. 
We do not take $\left( \frac{1}{ \sqrt{ 2 } } - s_{23} \right)$ as an expansion 
parameter because a rather large range is currently allowed and the situation 
will not be changed even with the next generation experiments \cite{MSS04}.

The $\epsilon$ perturbation theory based on (\ref{def-epsilon}) is a 
natural generalization of the framework taken by Cervera {\it et al.} 
\cite{golden} for systems with SI into the one with NSI. 
Note that the Cervera {\it et al.} formula 
(which we call the SI second-order formula) is the most widely used 
perturbative formula to discuss various aspects of neutrino oscillations. 
We derive the NSI second-order formula which generalizes the SI second-order formula to the case with NSI to have an overview of the neutrino oscillation 
phenomena in systems with NSI.
A different but related perturbative approach to neutrino oscillation with 
NSI has been studied in references \cite{ota1,blennow-etal,KTY-yasuda,kopp2,NSI-nufact,blennow-ohlsson,meloni-etal}.

In passing I have a few remarks on $\theta_{13}$. 
It is a big question whether $\theta_{13}$ falls into the range which can be 
explored by the next generation accelerator \cite{T2K,NOVA,europe} 
and the reactor \cite{reactor-exp} experiments. 
In fact, I argue sometimes rather strongly that $\theta_{13}$ must be large, 
for example in \cite{Nu2008-mina}.\footnote{
The basic reasoning for my belief is simple:\cite{NOVE06} 
The MNS matrix is the product of the two unitary matrices which diagonalize 
the neutrino and the charged lepton mass matrices. 
The two angles in the MNS matrix are known to be large. 
Then, why should the third one extremely small?
}
The belief is one of the motivations for my works which proposed reactor 
measurement of $\theta_{13}$ \cite{reactor-proposal} and superbeam 
measurement of lepton CP violation \cite{superbeam}. 
Nevertheless, I am a pessimist here with the ansatz (\ref{def-epsilon}). 
Well, the reason why I take the ansatz of small $\theta_{13}$ is that 
it is the only natural perturbative framework of neutrino oscillation. 
For instance, the appearance oscillation probability 
$P(\nu_e \rightarrow \nu_\mu)$ 
consists only of order $\epsilon^2$ terms. 
If I take a different ansatz 
$s_{13} \sim \sqrt{\epsilon} = \sqrt{ \frac{ \Delta m^2_{21} } { \Delta m^2_{31} } }$ 
(which roughly correspond to the Chooz limit \cite{CHOOZ}), 
the terms in $P(\nu_e \rightarrow \nu_\mu)$ 
do not scale uniformly and we would have to keep terms of order 
$\epsilon^{\frac{3}{2}}$ to include effects of CP violation. 
It necessitates to keep order $s^3_{13}$ terms.

\subsection{NSI second-order formula; $\nu_{e}$-related sector}

How can one go to the NSI second-order formula from the SI second-order formula? 
Though the task might look formidable, it is in fact trivial in $\nu_{e}$-related channels!
What is necessary is to make replacements in the atmospheric and the 
solar variables in the SI second-order formula and that's it: 
\begin{eqnarray}
s_{13} \frac{ \Delta m^2_{31} }{a} 
&\rightarrow& s_{13} \frac{ \Delta m^2_{31} }{a} + (s_{23} \varepsilon_{e \mu} + c_{23}  \varepsilon_{e \tau} ) e^{i \delta}, 
\nonumber \\ 
c_{12} s_{12}  \frac{ \Delta m^2_{21} }{a} 
&\rightarrow& c_{12} s_{12}  \frac{ \Delta m^2_{21} }{a}  + c_{23} \varepsilon_{e \mu} - s_{23} \varepsilon_{e \tau}.  
\label{replacement}
\end{eqnarray}
It is very easy to understand why the particular combinations of 
$\varepsilon$ parameters come in to the atmospheric and the 
solar variables, respectively. 
It is well known \cite{munich04} that in doing perturbation theory the 
convenient basis is the tilde basis 
$\tilde{H} = U_{23}^{\dagger} H U_{23} $. 
The combinations of the $\varepsilon$ parameters are the ones that appear 
in the NSI part of $\tilde{H}_{e 3}$ and $\tilde{H}_{e 2}$, analogues of 
``$\sin \theta_{13}$'' and ``$\sin \theta_{12}$''. 
(See equation (15) in \cite{NSI-perturbation}.)

The resultant NSI second-order formula in $\nu _e \rightarrow \nu _\mu$ 
channel reads \cite{NSI-perturbation}
\begin{eqnarray}
P(\nu _e \rightarrow \nu _\mu ) &=&
4 \Biggl 
 | \cn \biggl (\ci \si \frac{\mn}{a}
+\cn \varepsilon_{e\mu }-\sn \varepsilon_{e\tau }\biggr )
\sin \left( \frac{aL}{4E} \right) \exp{ \left( - i \frac{\mt L}{4E} \right) }
\nonumber \\
&& \hspace{-22mm} +
\sn \biggl (\st e^{-i\delta }\frac{\mt}{a}
+\sn \varepsilon_{e\mu }+\cn \varepsilon_{e\tau } \biggr )
\biggl (\frac{a}{\mt -a}\biggr )\sin \left( \frac{\mt -a}{4E}L \right)
\Biggr |^2.
\label{PemuB}
\end{eqnarray}
%
I hope the readers are convinced of my claim that 
the formula is surprisingly simple in its form. 
$P(\nu _e \rightarrow \nu _\tau )$ can be obtained by doing the transformation 
$c_{23} \rightarrow - s_{23}$ and $s_{23} \rightarrow c_{23}$ in 
$P(\nu _e \rightarrow \nu _\mu )$, but undoing any transformation in the 
generalized atmospheric and the solar variables defined in 
(\ref{replacement}).

A notable feature in (\ref{PemuB}) is that only the elements 
$\varepsilon_{e\mu }$ and $\varepsilon_{e\tau }$ appears in the 
NSI second order probability formula. 
Because of the decoupling of the other $\varepsilon$'s it is in principle 
possible to determine $\varepsilon_{e\mu }$ and $\varepsilon_{e\tau }$ 
together with the SI parameters $\theta_{13}$ and $\delta$, 6 real 
parameters including phases.
If one carries out this task by rate only analysis we need measurement of 
the oscillation probabilities in the following three channels
$\nu _e \rightarrow \nu _\mu$, $\nu _e \rightarrow \nu _\tau$, and 
$\nu _\mu \rightarrow \nu _e$  and their antineutrino counterpart.

\subsection{NSI second-order formula; $\nu_\mu - \nu_\tau$  sector}

In $\nu_\mu - \nu_\tau$ sector the situation is different. 
The NSI dependent piece in the oscillation probabilities 
$P(\nu _\mu \rightarrow \nu _\tau )$, 
$P(\nu _\mu \rightarrow \nu _\mu )$, and 
$P(\nu _\tau \rightarrow \nu _\tau )$ 
is universal. 
See \cite{NSI-perturbation} for explicit expressions. 
Because of this feature one cannot determine all the relevant NSI elements 
$\varepsilon_{\mu \tau}$ and $\varepsilon_{\mu \mu} - \varepsilon_{\tau \tau}$ 
(only the difference can be measured), 3 unknowns, 
by the rate only analysis.

The reasons for such curious feature of the universal NSI dependent term 
is simple to understand. By unitarity it follows that 
\begin{eqnarray}
P(\nu _\mu \rightarrow \nu _\mu) + P(\nu _\mu \rightarrow \nu _\tau) &=& 
1 - P(\nu _\mu \rightarrow \nu _e), 
\nonumber \\
P(\nu _\tau \rightarrow \nu _\tau) + P(\nu _\tau \rightarrow \nu _\mu) &=& 
1 - P(\nu _\tau \rightarrow \nu _e). 
\label{unitarity}
\end{eqnarray}
We note that $P(\nu _\mu \rightarrow \nu _e)$ and 
$P(\nu _\tau \rightarrow \nu _e)$ do not contain 
$\varepsilon_{\mu \mu }$, $\varepsilon_{\tau \tau }$, and 
$\varepsilon_{\mu \tau }$ to second order in $\epsilon$. 
Then, it follows from the first equation in (\ref{unitarity}) that 
$P(\nu _\mu \rightarrow \nu _\tau; \varepsilon_{\mu \mu }, \varepsilon_{\mu \tau }, \varepsilon_{\tau \tau }) = - 
P(\nu _\mu \rightarrow \nu _\mu; \varepsilon_{\mu \mu }, \varepsilon_{\mu \tau }, \varepsilon_{\tau \tau })$. 
Noticing that the terms related to 
$\varepsilon$'s in the $\nu_{\mu} - \nu_{\tau}$ sector are T-invariant,\footnote{
We emphasize that this feature itself is highly nontrivial, and can be realized 
only by an explicit computation.
}
the relations 
$P(\nu _\tau \rightarrow \nu _\tau; \varepsilon_{\mu \mu }, \varepsilon_{\mu \tau }, \varepsilon_{\tau \tau }) = - 
P(\nu _\mu \rightarrow \nu _\tau; \varepsilon_{\mu \mu }, \varepsilon_{\mu \tau }, \varepsilon_{\tau \tau }) $ must also hold. 
Therefore, the $\varepsilon_{\alpha \beta}$ ($\alpha, \beta = \mu, \tau$)
dependent term in the three channels are all equal up to the over-all sign. 
The necessity of spectrum analysis is obvious to determine all the NSI 
and the SI parameters.

\subsection{Summary table}

To answer to the questions raised above I present below 
the summary table. 
One of the features in Table.~\ref{order} which requires comment is that 
in the last column. 
In standard neutrino oscillation only with SI 
the matter effect comes in into the oscillation probability only at the 
second order in $\epsilon$, the property dubbed ``matter hesitation'' 
in \cite{NSI-perturbation}.  
It is the reason why it is so difficult to detect the matter effects in many 
accelerator neutrino experiments including NO$\nu$A \cite{NOVA}. 
The matter hesitation is a highly nontrivial property because we treat 
the coefficient $a$ (I mean, $a/\Delta m^2_{31}$) as of order unity. 
For example, there exists first order $a$ dependent term in the $S$ matrix, 
but it does not survive in $P(\nu_{e} \rightarrow \nu_e)$ 
because it enters as a phase factor. 
Notice, however, that its validity relies on the particular framework of 
perturbation theory. 
For a (simple!) proof of this property see \cite{NSI-perturbation}.

\begin{table}[h]
\vspace{- 1mm}
\caption{
Presented are the order in $\epsilon$ ($\sim 10^{-2}$) at which  
each type of $\varepsilon_{\alpha \beta}$ ($\alpha, \beta = e, \mu, \tau$) 
and  $a$ dependence 
($a$ is Wolfenstein's matter effect coefficient) 
starts to come in into the expression of the 
oscillation probability in $\epsilon$ perturbation theory.
The last column is for the $a$ dependence in the standard oscillation without NSI.
The order of $\epsilon$ indicated in parentheses implies the one 
for the maximal $\theta_{23}$ in which cancellation takes place in 
the leading order. 
}
\label{order}
\vspace{6mm}
 \small
\hspace{6mm}
  \begin{tabular}{lclclclclclclclclcl}\hline\hline
  { Channel } & $\varepsilon_{ee}$ & $\varepsilon_{e\mu}$ & $\varepsilon_{e\tau}$ & $\varepsilon_{\mu \tau}$ & $\varepsilon_{\mu \mu} $ & $\varepsilon_{\tau \tau}$ & a dep.(NSI) & a dep.(SI)  \\
  \hline
$P(\nu_{e} \rightarrow \nu_{\alpha})$: \\ $\alpha=e, \mu, \tau$ & $\epsilon^3$ & $\epsilon^2$ & $\epsilon^2$ &  $\epsilon^3$ & $\epsilon^3$ & $\epsilon^3$ & $\epsilon^2$ &$\epsilon^2$   \\  
  \hline
$P(\nu_{\alpha} \rightarrow \nu_{\beta})$: \\ $\alpha, \beta=\mu, \tau$ & $\epsilon^3$ & $\epsilon^2$ & $\epsilon^2$ &  $\epsilon^1 $ & $\epsilon^1 (\epsilon^2) $ & $\epsilon^1 (\epsilon^2) $ & $\epsilon^1$ &$\epsilon^2$   \\  
  \hline\hline
\end{tabular} 
\vspace{-5mm}
\end{table}
\vspace{6mm}

One of the implication of the matter hesitation is that $\varepsilon_{e e}$ 
comes into the oscillation probability at order $\epsilon^3$ in all channels, 
as indicated in Table.~\ref{order}. It is because $\varepsilon_{e e}$ is 
nothing but a small shift of the matter effect coefficient $a$. 
Because of this property it is very difficult to measure $\varepsilon_{e e}$ 
in long-baseline experiments. 
It should be remarked, however, that assuming that the other NSI elements 
are vanishingly small it can be measured in a great precision of a few \% 
at a neutrino factory \cite{mina-uchi,gandhi-winter}. But, I must note that 
it is only true under the assumption that the earth matter density along 
the neutrino trajectory is accurately known.

Another notable feature in Table.~\ref{order} is that there exists first order 
term of NSI element $\varepsilon_{\mu \tau}$ 
(and $\varepsilon_{\mu \mu} - \varepsilon_{\tau \tau}$ if 
$\theta_{23} \neq \frac{\pi}{4}$) in the $\nu_\mu - \nu_\tau$ sector. 
Clearly, they are due to direct transition caused by these NSI elements. 
In fact, rather high sensitivities for determining 
$\varepsilon_{\mu \tau}$ and 
$\varepsilon_{\mu \mu} - \varepsilon_{\tau \tau}$ observed 
in atmospheric \cite{fornengo} and future accelerator \cite{NSP-T2KK} 
neutrino analyses are benefited by this feature.

\subsection{SI$-$NSI confusion}

The structure of the oscillation probability (\ref{PemuB}) 
in which $\varepsilon_{e \alpha}$ enters into the expression 
only through the generalized solar and the atmospheric variables 
(\ref{replacement}) is one of the most significant features of the 
oscillation probability with NSI. 
It also implies that, when the parameter determination is attempted, 
there exists severe confusion between the SI and the NSI parameters.  
Hence, our result may be regarded as an analytic proof of 
the general ``confusion theorem'' in $\epsilon$ perturbation theory.

The uncovered structure may be helpful to formulate a strategy of resolving 
the confusion, because (\ref{replacement}) clearly dictates which SI parameters 
will be confused by which NSI variables by which way. 
Notice that our confusion theorem is quite different in nature from the one 
proved in \cite{confusion2} in which $\theta_{13}$ is confused with 
the NSI elements in production and detection processes.

\subsection{Parameter degeneracy in neutrino oscillation with NSI}

It is now well understood that phenomenon of parameter degeneracy, 
existence of the multiple solutions, occurs in neutrino oscillation 
measurement of SI parameters 
\cite{intrinsic,MNjhep01,octant}. 
Because of the large number of  unknown (i.e., to be determined) parameters 
(2 standard and 8 NSI ones) the parameter degeneracy in the full system 
is a formidable problem to work out, even under the approximation of ignoring 
NSI effects in production and detection of neutrinos. 

Yet, it was possible to recognize a completely new type of degeneracy 
\cite{NSI-perturbation}. 
Let us denote the generalized atmospheric and the solar variables in (\ref{replacement}) in section~5.2 as $\Theta_{\pm}$ and $\Xi e^{-i \delta}$, respectively. 
Then, if there is a solution 
$\vert \Theta_{\pm}^{(1)} \vert$ and $\vert \Xi^{(1)} \vert$, then 
the second solution 
$\vert \Theta_{\pm}^{(2)} \vert = \sqrt{  \frac{ Z }{ X_{\pm} } } \vert \Xi^{(1)} \vert$ and 
$\vert \Xi^{(2)} \vert =  \sqrt{  \frac{ X_{\pm} }{ Z } }  \vert \Theta_{\pm}^{(1)} \vert$  
exists. 
(See \cite{NSI-perturbation} for definitions of $X_{\pm}$ etc.) 
It can be called the ``atmospheric$-$solar variables exchange'' degeneracy, 
which arises because of large number of unknown parameters in the solar 
and the atmospheric variables. 
Of course, it does not survive when NSI is turned off because there is no 
solar degrees of freedom (as to be determined parameters) as can be 
seen in (\ref{replacement}).

What is the right way in this difficult problem of degeneracy in systems with NSI? 
As a first step, we have worked out the problem in a region where 
the matter effect can be treated as a perturbation. 
For early references of matter perturbation theory, see e.g., \cite{AKS,MNprd98}. 
It is known that analysis of the parameter degeneracy becomes particularly transparent in this setting \cite{MNjhep01,resolve23,T2KK1st,T2KK2nd}.

Our analysis is most transparent in the ``discrete'' degeneracy, the 
sign-$\Delta m^2_{31}$ and the octant ones. 
Let me describe first the sign-$\Delta m^2_{31}$ degeneracy. 
One can show that the NSI dependent terms in the oscillation probability 
$P(\nu _e \rightarrow \nu _\alpha )$ ($\alpha = \mu, \tau$) 
to first order in $a$ is invariant under the transformation 
\begin{eqnarray}
\Delta m^2_{31} &\rightarrow& - \Delta m^2_{31}, 
\nonumber \\ 
\delta &\rightarrow& \pi - \delta, 
\nonumber \\ 
\phi_{e \alpha} &\rightarrow& 2\pi - \phi_{e \alpha}, 
\label{extended-symmetry}
\end{eqnarray}
while keeping $\theta_{13}$ and $\vert \varepsilon_{e \alpha} \vert$ 
fixed. It nicely complements the discussion in \cite{MNjhep01} and 
it indicates that there exists a new (approximate) solution 
with differing sign of $\Delta m^2_{31}$. 
It is worth to note that the invariance is true only if the CP phase of 
NSI element is involved in the transformation (\ref{extended-symmetry}).

Similarly, one can show that there is another invariance under the transformation 
(assuming $\theta_{23} \neq \frac{\pi}{4}$) 
\begin{eqnarray}
c_{23} &\rightarrow& s_{23}, 
\nonumber \\ 
s_{23} &\rightarrow& c_{23}, 
\nonumber \\ 
(\varepsilon_{\mu \mu }-\varepsilon_{\tau \tau }) &\rightarrow& 
- (\varepsilon_{\mu \mu }-\varepsilon_{\tau \tau }). 
\label{extended-symmetry2} 
\end{eqnarray}
It means that the $\theta_{23}$ octant degeneracy prevails in the 
presence of NSI, and actually in an extended form which involves 
NSI parameter $\varepsilon_{\mu \mu }-\varepsilon_{\tau \tau }$. 
Since this NSI parameter decouples from $P( \nu_{\mu} \to \nu_{e} )$ 
to second-order in $\epsilon$, the presence of the $\theta_{23}$ 
octant degeneracy remains intact when the NSI is included though values 
of the degenerate solutions themselves are affected by the presence of 
$\varepsilon_{e \alpha }$. 
Thus, we have shown that the parameter degeneracy survives the presence 
of NSI provided that NSI elements are ``actively involved'' in the 
degenerate solutions.

The other salient feature of degeneracy in the mass perturbative regime 
is the property called the ``decoupling between degeneracies'' \cite{T2KK2nd}. 
We have revisited this issue in the systems with and without NSI.
The conclusion obtained in \cite{NSI-perturbation} is that the decoupling 
between the sign-$\Delta m^2_{31}$ and the $\theta_{23}$ octant 
degeneracies holds with and without NSI. 
On the other hand, decoupling between them and the intrinsic degeneracy 
does not holds with and without NSI, partly correcting the conclusion in \cite{T2KK2nd}. 
The only exception is the special setting at the oscillation maximum, 
or more precisely, the shrunk ellipse limit \cite{KMN02}, for which 
the decoupling holds.

\subsection{Parameter degeneracy; An example}

\begin{figure}[h]
\vspace*{-0.2cm}
\begin{center}
\epsfig{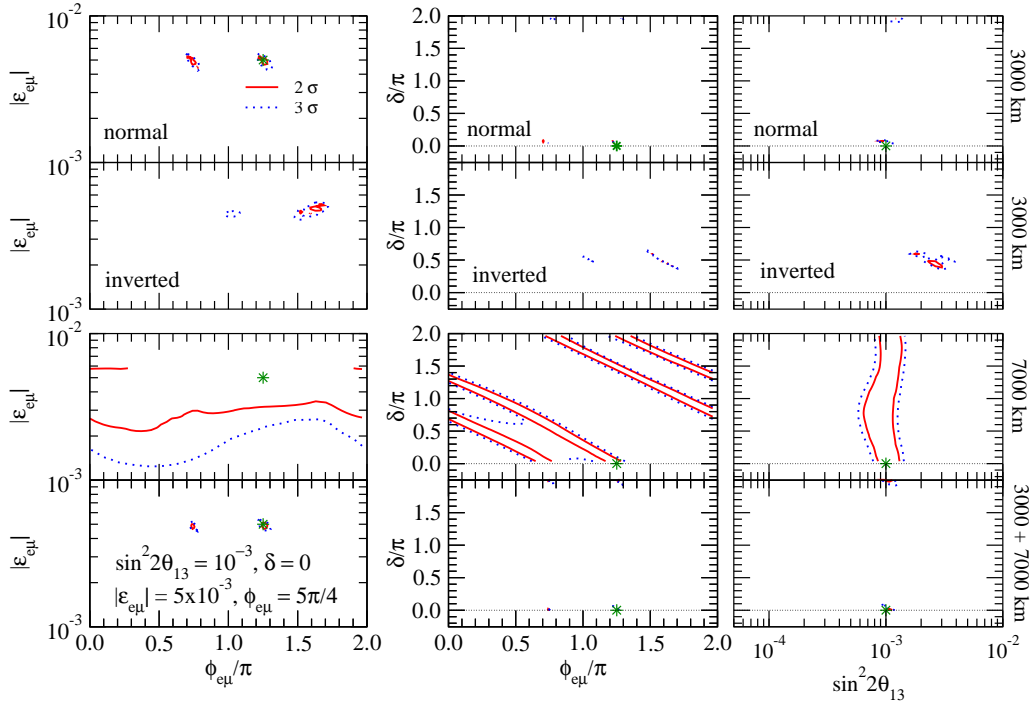} 
\end{center}
\caption{
Allowed regions in the $\phi_{e\mu}-|\varepsilon_{e\mu}|$
  plane (left column), $\phi_{e\mu}-\delta$ plane (middle column) and
  $\sin^2 2\theta_{13}-\delta$ plane (right column) corresponding to 2
  and 3 $\sigma$ CL obtained for the system with $\varepsilon_{e\mu}$.
  Panels in the upper 2 rows (3rd row) correspond to the case where
  only a single detector at 3000 km (7000 km) is taken into account,
  whereas the ones in the 4th row correspond to the case where results
  from the two detectors are combined.  The input parameters are taken
  as: $\sin^2 2\theta_{13} = 10^{-4}$, $\delta=0$,
  $|\varepsilon_{e\mu}|= 5 \times 10^{-3}$ and $\phi_{e \mu}=5\pi/4$
  (indicated by the green asterisk), and the mass
  hierarchy is normal.  For the case where only the single detector at 3000 km
  is taken into account, allowed regions exist not only for the normal
  mass hierarchy regime but also for the inverted one, as shown in the panels in 
  the second row. 
}
\label{para-dege-emu}
\end{figure}

An example of the parameter degeneracy is presented in Fig.~\ref{para-dege-emu}. 
This is one of the examples found in doing the work \cite{NSI-2phase} 
but is not presented in the reference. 
It clearly demonstrates the existence if the $\Delta m^2_{31}$-sign flipped 
and the intrinsic degeneracies, the natural extension of the one 
\cite{intrinsic,MNjhep01} to the systems with NSI.
As pointed out in section 5.6 the phase $\phi_{e\mu}$ of 
the NSI element is indeed heavily involved \cite{NSI-perturbation} though 
the relation (\ref{extended-symmetry}) which is valid in the mass perturbative 
regime does not quite hold.
In this example (as well as in the one which is presented in \cite{NSI-2phase}), 
the far detector measurement successfully lift the $\Delta m^2_{31}$-sign 
flipped degeneracy, but not and the intrinsic one.

\begin{figure}[h]
\vspace*{3mm}
\begin{center}
\epsfig{figure=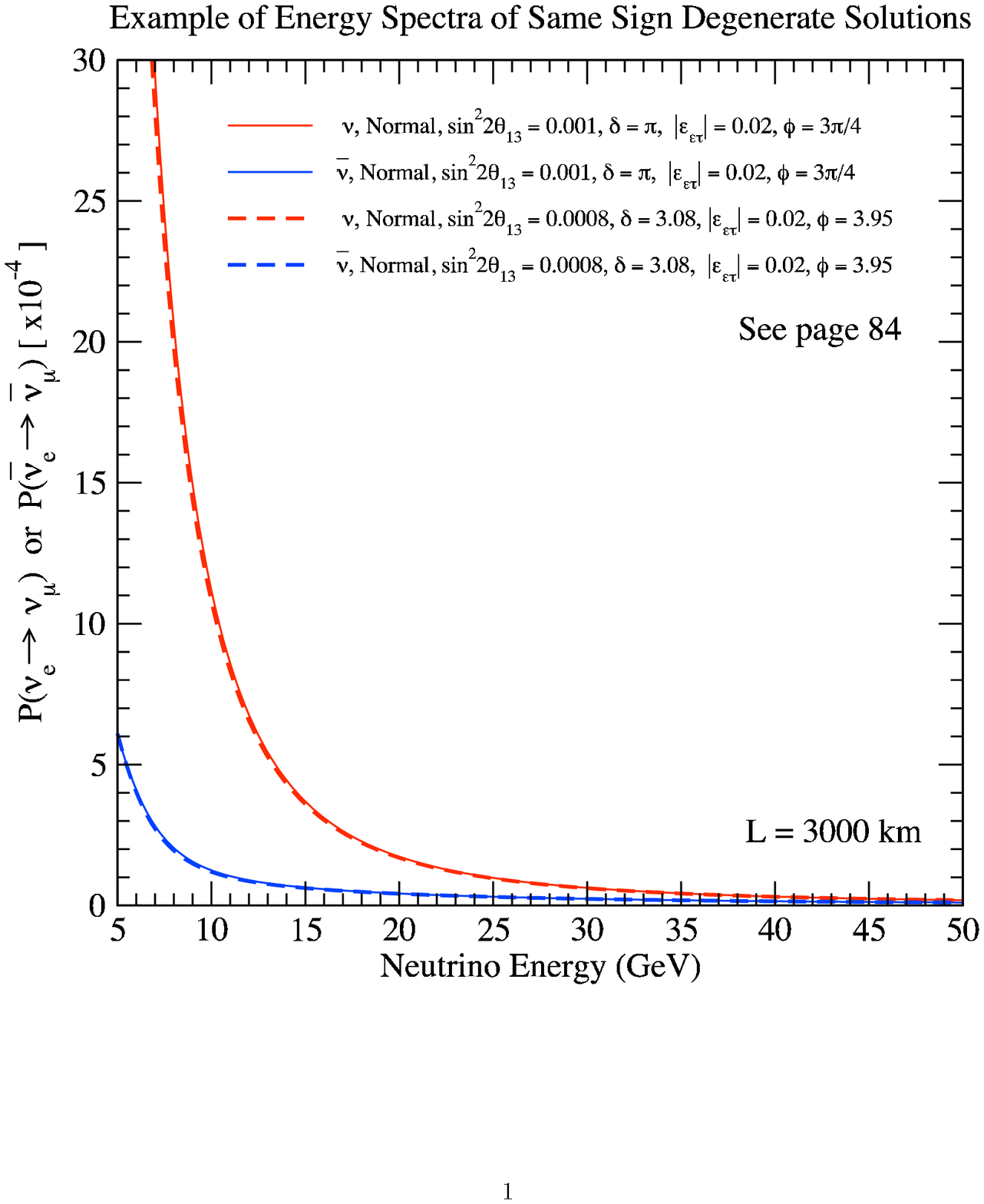,width=64mm} 
\hspace{2mm}
\epsfig{figure=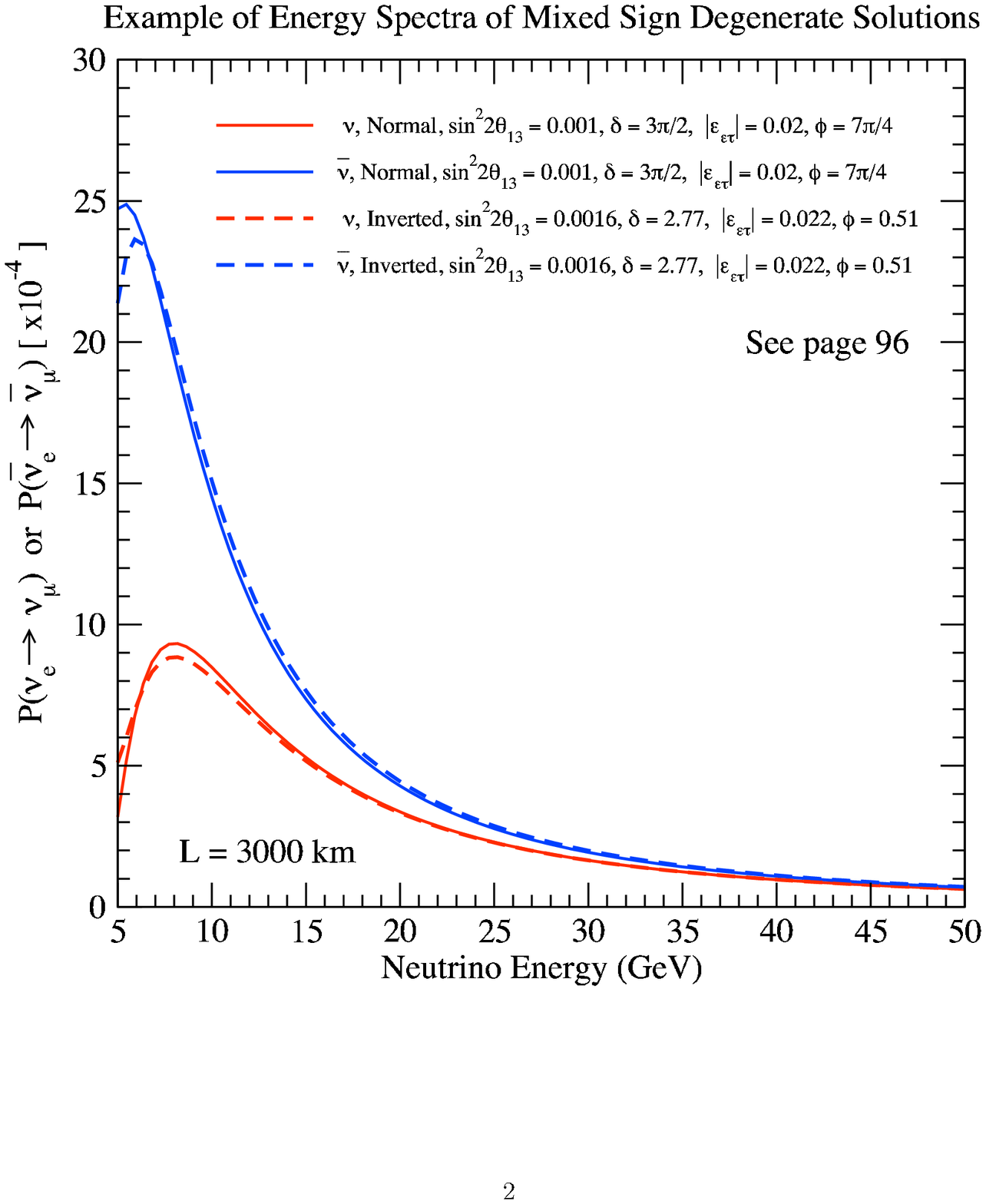,width=64mm} 
\end{center}
\vspace*{-10mm}
\caption{
Energy spectra of the oscillation probability for a
  system with non-zero NSI element $\varepsilon_{e \tau}$ corresponding to the
  same $\Delta m^2_{31}$-sign intrinsic degeneracy (left panel) and
  the flipped $\Delta m^2_{31}$-sign degeneracy (right panel). 
}
\label{energy-spectra-etau}
\end{figure}

How robust is the degeneracy in system with NSI?
It is a difficult question to answer in general. 
But, it is worth to remark that sometimes the degeneracy is extremely hard to 
solve because the energy spectra corresponding the degenerate solutions 
are so similar.\footnote{
It is often the case that in systems only with SI that the intrinsic degeneracy 
can be ``easily'' resolved by the spectrum analysis. 
See for example, in the case of T2K or T2KK settings \cite{T2KK1st,T2KK2nd}. 
}
%
This fact is demonstrated in Fig.~\ref{energy-spectra-etau} in the 
$\varepsilon_{e\tau}$ system with the particular values 
of the SI and NSI parameters. 
See Fig.~13 in \cite{NSI-2phase} for the corresponding figures for the 
$\varepsilon_{e\mu}$ system with exactly the same feature.

\section{Discriminating CP violation due to SI and NSI phases}

I emphasize that one of the most important features of the system with 
NSI is the coexistence of two kind of CP violation \cite{concha1}, the one 
due to $\delta$ in the MNS matrix \cite{MNS}, the leptonic version of 
the celebrated Kobayashi-Maskawa (KM) phase \cite{KM} in the CKM matrix \cite{KM,cabibbo} for quarks, 
and the other which come from the phases of NSI elements.
Knowing the nature of CP violation seen in any kind of experiments \cite{CPV-review} 
is of decisive importance because of many reasons, in particular for possible 
connection to leptogenesis scenario \cite{leptogenesis}, currently 
the most promising one for baryon number generation in the universe.

\subsection{Two-detector setting}

I have discussed in the last year in Venice the possibility of resolving the 
$\theta_{13} -$NSI confusion \cite{confusion1,confusion2} 
by the near (3000 km)$-$far (7000 km) two-detector setting in neutrino 
factory based on our work \cite{NSI-nufact}. 
I use the same setting to examine the question of whether the two-phase 
confusion can be resolved \cite{NSI-2phase}. 
For a related work on the same subject see 
\cite{winter-nonstandardCP}.
The similar question of distinguishing two kind of CP violation in the 
context of ``unitarity violation'' approach \cite{unitarity-violation} 
has also been investigated in \cite{altarelli-meloni,antusch-etal}.

Setting of the second detector at around the magic baseline $\simeq 7000$ km 
was motivated by high sensitivity to the matter effect \cite{mina-uchi}.
It is in concordant with the similar two detector setting in a neutrino factory 
as a degeneracy solver \cite{intrinsic,huber-winter,ISS-report}. 
Two-detector setting has been proposed in neutrino experiments 
in a variety of contexts.\footnote{
Here is a brief history of two-detector settings in contemporary neutrino 
experiments: 
It was proposed as an appropriate setting for measuring CP 
violation \cite{MNplb97} in the context of low energy superbeam 
experiment \cite{superbeam}. 
In a quite different context of reactor measurement of $\theta_{13}$
\cite {reactor-proposal} the two-detector setting is the standard one to 
guarantee the near-far cancellation of systematic errors. 
It has triggered interests in the world wide scale \cite{reactor-white}, 
and led to the several international collaboration experiments \cite {reactor-exp}. 
The idea has also been applied to the Kamioka-Korea two detector complex 
with an upgraded neutrino beam from J-PARC to determine the mass hierarchy 
as well as discovering CP violation \cite{T2KK1st,T2KK2nd}. 
It ``unifies'' the two aspects of near-far cancellation and synergy between 
the two detectors, and can serve for a possible upgrade option of the 
T2K experiment \cite{T2K}. 
For an overview of T2KK, see e.g., \cite{T2KK-review}, and 
for a review of the two-detector setup \cite{ICFP07-mina}. 
}
%
The basic idea in the present context is to seek complementary role 
played by the far detector.

\subsection{Use of the bi-probability plot}

Though I told you that derivation of the NSI second-order formula 
of the oscillation probability is simple, it means neither that the dynamics 
of the system is simple, nor it is easy to understand. 
What makes the system so complicated is the very existence of two CP 
violating phases, $\delta$ and the phase $\phi_{e\alpha}$ of the NSI element 
$\varepsilon_{e\alpha} = \vert  \varepsilon_{e\alpha}  \vert e^{ i \phi_{e\alpha} } $ ($\alpha = \mu, \tau$). 
In my talk, therefore, I report the work done with only a single type of NSI, 
either $\varepsilon_{e\mu}$ or $\varepsilon_{e\tau}$, as a first step of 
understanding the features of neutrino oscillation with NSI.

\begin{figure}[h]
\vspace*{-0.2cm}
\begin{center}
\epsfig{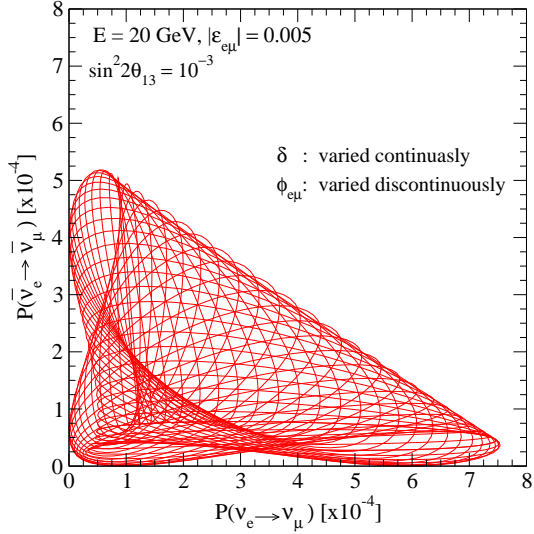} 
\end{center}
\caption{
Bi-probability plots in 
$P(\nu_e \to \nu_{\mu}) - P(\bar{\nu}_e \to \bar{\nu}_{\mu})$ space 
at $L=3000$ km, for $E=$ 20 GeV, with $\sin^2 2\theta_{13}=10^{-3}$ 
and $\varepsilon_{e\mu}=5 \times 10^{-3}$ 
computed numerically using the constant matter density $\rho =  3.6$ g/cm$^3$ 
assuming the electron number density per nucleon of 0.5.  
The both axes is labeled in units of $10^{-4}$. 
The values of the parameters taken are 
$\varepsilon_{e \mu}$
$\sin^2 2\theta_{13}$ 
}
\label{byprob-20GeV}
\end{figure}

How complicated is the system with NSI?
Seeing is believing. 
Presented in Fig.~\ref{byprob-20GeV} is the bi-probability plot in 
$P( \nu_{e} \rightarrow \nu_{\mu} ) - P( \bar{\nu}_{e} \rightarrow \bar{\nu}_{\mu} )$ space \cite{MNjhep01} but by varying the two phases, $\delta$ and $\phi_{e\mu}$.
As you see the ellipses move around in the plane such that the whole 
triangular region is (almost) swept over. 
So ``anything can happen'' with the two phases. 
We have characterized the behavior of ellipses as {\em rotating ellipses} in \cite{NSI-2phase}; 
An ellipse drawn by varying the phase A rotates when the other phase B is varied. 
Because of the behavior of the probabilities rich phenomena such as confusion between SI and NSI parameters and the parameter degeneracy are expected.

%
\vspace*{2mm}
%
\begin{figure}[h]
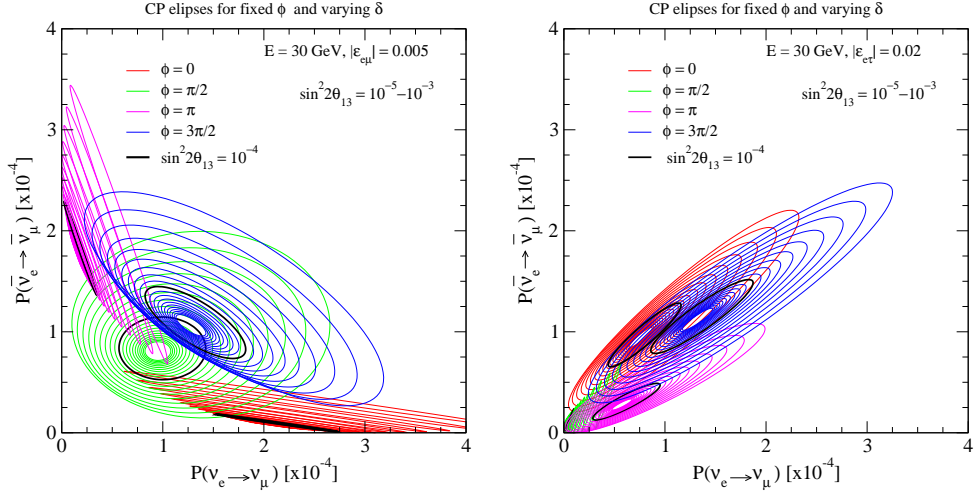

\vspace*{2mm}
\begin{center}
\epsfig{figure=Prob_vary_delta_theta13_emu.eps,width=62mm} 
\hspace{2mm}
\epsfig{figure=Prob_vary_delta_theta13_etau.eps,width=62mm} 
\end{center}
\caption{
Bi-probability plots drawn by continuously varying $\delta$ 
  for four different values of $\phi$.  
}
\label{biP-emu-etau}
\end{figure}

In fact, various viewpoints have to be involved to really understand features 
of neutrino oscillations with NSI and the sensitivities to the NSI elements, 
$\vert \varepsilon_{e \alpha} \vert$ and $\phi_{e \alpha}$ ($\alpha=\mu, \tau$), 
 and the SI parameters, $\delta$ and $\sin^2 2\theta_{13}$, to be achieved 
 by the detectors at $L=3000$ km and $L=7000$ km separately and in combination. 
They include:  
\begin{itemize}

\item 
How prominent is the synergy between the near and the far detectors 
for determination of SI and NSI parameters?
How it differs between the systems with $\varepsilon_{e \mu}$ and  
$\varepsilon_{e \tau}$? What about dependence on values of the parameters, 
in particular on the size of NSI elements? 

\item 
How can the two-phase confusion be resolved?
Does the answer depend on which NSI element is turned on?

\item 
What is the nature of the parameter degeneracy in system with NSI, 
and whether it can be resolved by the two-detector setting? 
If so how can it be realized?

\end{itemize}

\noindent
These points are fully discussed in our paper \cite{NSI-2phase}. 

Here, we make comments only on a puzzle. 
The difference in sensitivity to NSI has non-trivial features. 
At relatively large values of NSI, $\varepsilon_{e \mu}=10^{-3}$ and 
$\varepsilon_{e \tau}=10^{-2}$, the size of the ellipses are similar in size. 
But, the sensitivity is in fact very different between the two systems; 
The one in the $\varepsilon_{e \mu}$ system is much higher than that in the 
$\varepsilon_{e \tau}$ system. 
On the other hand, the parameter degeneracy is much severer at the
near detector in systems with $\varepsilon_{e\mu}$ compared to the
ones with $\varepsilon_{e\tau}$ as is shown in the tables.

A unified understanding of the puzzling features becomes possible once
one draws the bi-probability plots by varying $\delta$ for several different 
values of $\phi$ and $\theta_{13}$, as done in Fig.~\ref{biP-emu-etau}. 
The degeneracy is severer in the $\varepsilon_{e\mu}$ system because
of the more dynamic behavior of the bi-probability ellipses
as shown in the left panel of Fig.~\ref{biP-emu-etau}.
Because the ellipses can locate themselves essentially everywhere in the
bi-probability space there are chances that fake solutions can be
produced at points far apart from the true solution.
What about the difference in the sensitivities?
We observe in the right panel of Fig.~\ref{biP-emu-etau} 
that the ellipses in the $\varepsilon_{e\tau}$ system remain in 
much more compact region when $\delta$ and $\theta_{13}$ are varied.
Because of the finite resolution of the experimental data it appears that the
dense concentration of the ellipses with different parameters
leads to merging of many degenerate solutions. 
It probably explains lack of the sensitivities and at the same time much 
less frequent degenerate solutions in the $\varepsilon_{e\tau}$ system.

\section{Discovery Potentials}

This is the appropriate point to discuss the discovery potential of 
various quantities, 
$\vert \varepsilon_{e \mu} \vert$, 
$\vert \varepsilon_{e \tau} \vert$, 
$\phi_{e \mu}$, $\phi_{e \tau}$, 
as well as the standard parameters $\delta$ (and $\theta_{13}$ in principle) 
and the neutrino mass hierarchy. 
In this report we focus on CP violation caused by the lepton KM 
phase $\delta$, and the phase $\phi$ of NSI. 
The discovery potential for the rest of the quantities are discussed in 
\cite{NSI-2phase}.

By the way, I remind you that all the figures presented in this manuscript 
are new, i.e., no single figure which is identical to the one in \cite{NSI-2phase}. 
To keep this tradition I will always give in this manuscript the sensitivity 
regions calculated with the inverted mass hierarchy as input. 
Notice that all the sensitivity plots given in \cite{NSI-2phase} are calculated 
by taking the normal hierarchy as input.
Great thanks to Hiroshi Nunokawa for his efforts to prepare them for this manuscript.

While I do not give any details of the quantitative analysis in this 
manuscript (for which see \cite{NSI-2phase}), 
it should be remarked that all the systematic errors 
as well as backgrounds are ignored in our analysis. 
Here, I explain the reasons for this choice in my own language.
The dimension six operator that can give neutrino's NSI without producing 
unwanted four charged lepton NSI is unique \cite{gavela-etal}, 
the anti-symmetrized one given in Eq.~(\ref{dim6}) in section 3. 
Then, unless someone is able to show that {\em only the} dimension six 
operator naturally arises in a certain class of models of new physics at 
TeV scale,\footnote{
From the reasoning below, I think it important to pursue this possibility.
}
we must prepare for search for the dimension eight (or higher) 
operators, with the size $\vert \varepsilon \vert \sim 10^{-4}$ 
assuming no extra suppression.\footnote{
I have assumed throughout this report that discussions on NSI in the lepton 
sector applies to the operators which involve neutrinos and quarks. 
Though I think it reasonable it can be subject to criticism. 
}
%
If it turned out to be the case one must think of the experimental technology 
which can accommodate this request. 
Since its realization is not known, we invented a model experiment using 
the neutrino factory setting with an ideal detector for which the 
systematic errors and background are ignored.

\subsection{CP violation due to NSI}

\begin{figure}[h]
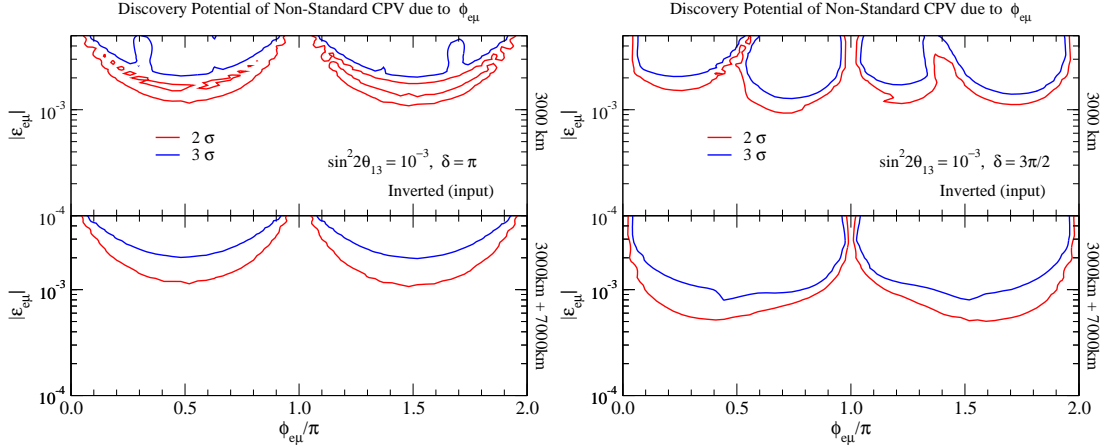

\vspace*{3mm}
\begin{center}
\epsfig{figure=NSI_NonStandard-CPV_pi_em_inv.eps,width=72mm} 
\epsfig{figure=NSI_NonStandard-CPV_3piby2_em_inv.eps,width=72mm} 
\end{center}
\caption{
Regions where the non-standard CP violation caused by $\phi_{e \mu} \ne 0$
  or $\phi_{e \mu} \ne \pi$ can be established for the case $\sin^2
  2\theta_{13} = 10^{-3}$, $\delta = \pi$ (left panel) and $\delta =
  3\pi/2$ (right panel). 
  The inverted mass hierarchy is assumed as the input.  }
\label{NSI-CPV-sensitivity-em}
\end{figure}

\begin{figure}[h]
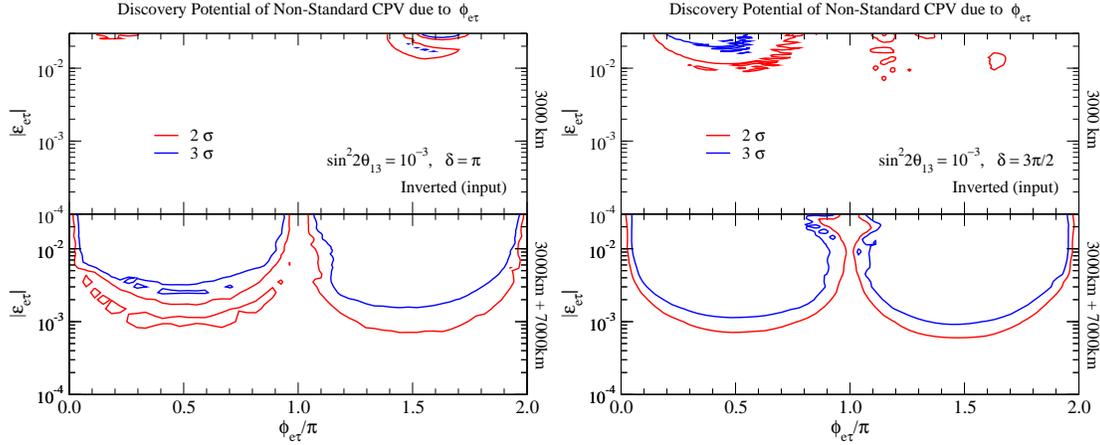

\vspace*{3mm}
\begin{center}
\epsfig{figure=NSI_NonStandard-CPV_pi_et_inv.eps,width=72mm} 
\epsfig{figure=NSI_NonStandard-CPV_3piby2_et_inv.eps,width=72mm} 
\end{center}
\caption{
The same as in Fig.~\ref{NSI-CPV-sensitivity-em} but for the non-standard 
CP violation caused by $\phi_{e \tau} \ne 0$ or $\phi_{e \tau} \ne \pi$.  
}
\label{NSI-CPV-sensitivity-et}
\end{figure}

Let us start with the sensitivity to CP violation caused by the phase $\phi$ of NSI.
In Fig.~\ref{NSI-CPV-sensitivity-em} and Fig.~\ref{NSI-CPV-sensitivity-et}
the regions sensitive to non-standard CP violation due to NSI are presented 
in $\phi - \vert \varepsilon \vert$ space. 
I these regions one can detect non-standard CP violation by NSI 
($\phi \neq 0$ and $\phi \neq \pi$) at 2$\sigma$ (red thin lines) and 
3$\sigma$ (blue thick lines) CL. 
Fig.~\ref{NSI-CPV-sensitivity-em} and Fig.~\ref{NSI-CPV-sensitivity-et} are for 
the $\varepsilon_{e\mu}$ and the $\varepsilon_{e\tau}$ systems, respectively.

%

By comparing the Fig.~\ref{NSI-CPV-sensitivity-em} and 
Fig.~\ref{NSI-CPV-sensitivity-et} to Figs.~18 and 20 in \cite{NSI-2phase}, 
respectively, 
one notices several notable differences between the normal and the 
inverted mass hierarchies. 
In the $\varepsilon_{e\mu}$ system the sensitivities to non-standard CP 
violation at $\delta=\pi$ are significantly worse both at the near detector 
(3000 km) and the near-far (7000 km) combined in comparison to 
those obtained with the input normal hierarchy.
The results at $\delta=3\pi/2$, however, are very similar to the case of 
normal mass hierarchy. 
A somewhat curious behavior seen in the upper-right panel of 
Fig.~\ref{NSI-CPV-sensitivity-em}, no sensitivity to non-standard CP violation 
at the maximal CP violating inputs, $\phi = \pi/2$ and $\phi = 3\pi/2$ is 
explained as a consequence of the parameter degeneracy, the one 
called the $\phi$ degeneracy in \cite{NSI-2phase}.

In the $\varepsilon_{e\tau}$ system the sensitivities to non-standard CP 
violation are similar to the normal hierarchy case. 
The most notable difference is in the $\delta=\pi$ case; 
At the near detector the sensitivity to non-standard CP violation is a 
bit worse than that of the normal hierarchy case, but curiously enough 
it is a little better when the far detector is combined. 
It can well be the case because the features of synergy between the two 
detectors are highly nontrivial \cite{NSI-nufact,NSI-2phase}.

\subsection{Standard CP violation}

\begin{figure}[h]
\vspace*{2mm}
\begin{center}
\epsfig{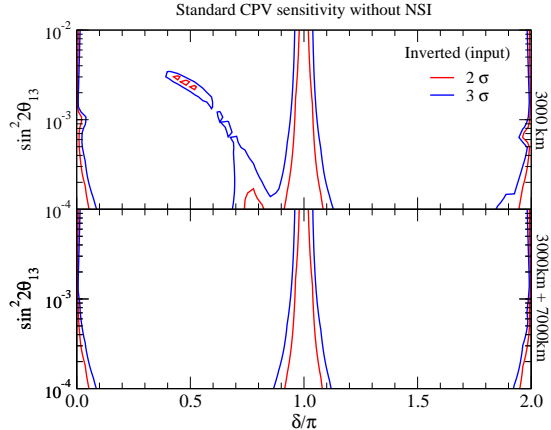} 
\end{center}
\vspace*{-2mm}
\caption{
Sensitivity to discovery of standard CP violation. Here 
no effect of NSI is assumed in the input data or in the fit. 
The upper panel shows the case where only the detector at 3000 km is 
considered, whereas the lower panel is the case corresponding
to the combination of detectors at two different baselines. 
The inverted mass hierarchy is assumed as input. 
}
\label{cpv-sensitivity-noNSI}
\vspace*{-6mm}
\end{figure}

Let us go back to the sensitivity to CP violation due to the KM phase $\delta$.
In Fig.~\ref{cpv-sensitivity-noNSI} the regions sensitive to the standard 
CP violation due to $\delta$ are presented in the system without NSI. 
They are significantly worse than the normal hierarchy case given in 
Fig.~22 in \cite{NSI-2phase}. 
Most probably, it is due to relatively smaller number of events in the 
antineutrino channel. 
Notably a peninsula like region without sensitivity develops from 
$\delta \simeq 0.8 \pi$ to $\delta \simeq 0.5 \pi$.

\begin{figure}[h]
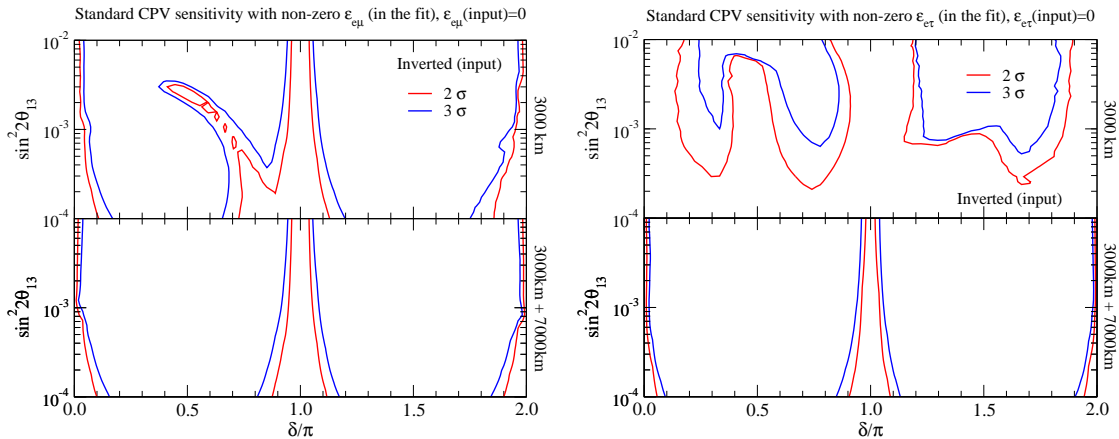

\begin{center}
\epsfig{figure=CPV_NSI_em_inv.eps,width=72mm} 
\hspace{1mm}
\epsfig{figure=CPV_NSI_et_inv.eps,width=72mm} 
\end{center}
\caption{
The similar plots as in Fig.~\ref{cpv-sensitivity-noNSI} but 
with non-zero NSI allowed in the fit; 
The input data was generated assuming the inverted mass hierarchy 
without NSI but non-zero values of $\varepsilon_{e\mu}$ (left panel) 
and $\varepsilon_{e\tau}$ (right panel) were allowed in the fit. 
}
\label{cpv-sensitivity-NSI}
\end{figure}

The gross features of the sensitivity regions remain unchanged even when 
the NSI degrees of freedom is turned on, as can be seen in 
Fig.~\ref{cpv-sensitivity-NSI}. 
The sensitivities to the standard CP violation are slightly worse 
compared to the normal hierarchy case given in Fig.~23 in \cite{NSI-2phase}.

I must warn the readers that the sensitivity contours are unstable to 
inclusion of the systematic errors and backgrounds. 
Yet, I suspect that these sensitivity regions are similar to the ones obtained 
with the apparatus which can explore the NSI in the whole region down to 
$\vert \varepsilon \vert \sim 10^{-4}$, the sensitivity which I argued to be 
necessary in the future NSI search.

%

\section{Conclusion}

After reviewing the theoretical progresses on NSI recently made in section~3, 
I tried to explain the works done by our group on theoretical and 
phenomenological aspects of hunting the NSI in sections 5-7. 
I guess the former contributed to illuminate the global features of neutrino 
oscillation with NSI including method for parameter determination and 
recognition of the parameter degeneracy with NSI. 
While in the latter we have investigated the problem of discriminating the 
two kind of CP violation, one due to the standard KM phase and the other 
by phases of NSI elements.

It appears that the near (3000 km) $-$ far (7000 km) two detector setting in 
neutrino factory does have a  rather high sensitivity to explore 
$\vert \varepsilon_{e\mu} \vert$ to 
$\simeq 10^{-4}$ in a lucky region of $\phi$ and to 
$\simeq$ several $\times 10^{-4}$ in most of the region of $\phi$. 
The sensitivity to $\vert \varepsilon_{e\tau} \vert$ is lower but still it can be 
explored to $\simeq 10^{-3}$. 
See Figs.~14-17 in \cite{NSI-2phase}. 
The sensitivity to CP violation is also very good; 
The one due to NSI phase can be probed to 
$\vert \varepsilon_{e\mu} \vert$ to $\simeq$ several $\times 10^{-4}$ in most 
of the region of $\delta$ and $\phi_{e\mu}$. 
They are close but not quite the lower end of the required full region for 
exploration  of NSI due to TeV scale new physics. 
Moreover, our estimation ignores the systematic errors and backgrounds, 
and hence is overly optimistic one.

How can this situation be overcome? Honestly, I don't know the answer. 
However, a few comments may be made:  
\begin{itemize}

\item 
Effects of NSI in production and detection of neutrinos, which are completely 
ignored in our works, can be of great help. 

\item 
One can formulate the good enough reasoning to convince people that 
search for neutrino's NSI to $\vert \varepsilon \vert \sim10^{-2}$ is 
sufficiently informative to signal new physics at TeV scale.

\end{itemize}

\noindent
However, I guess the former possibility is not easy to be realized.
Nonetheless, I would like to recall that no discovery done in the past was 
an easy one. 
Also, upon identification or grasp of the new physics we should obtain the 
clearer view. 
A second eel can be a big one!

\section{Bibliographical Note}

Given the large number of references devoted to the subject of NSI 
it is not easy to find the appropriate one, in particular for a newcomer 
as I was sometime ago. 
Therefore, I tried some efforts to collect them here with classification by subjects. 
Yet, it is extremely difficult to find all of them, and therefore, the list should 
be considered as an incomplete one. 
I would like to apologize to those who will find their references missed.
The categories I use are: 

\begin{itemize}

\item 
Accelerator neutrinos; neutrino factory 
\cite{confusion1,confusion2,ota1,NSI-nufact,NSI-2phase,NSI-accelerator-nufact}.

\item 
Accelerator neutrinos; excluding neutrino factory \cite{kopp2,NSP-T2KK,NSI-accelerator-not-nufact}.

\item 
Atmospheric neutrinos \cite{concha1,fornengo,NSI-atmospheric}.

\item 
Reactor or spallation source neutrinos or low-energy scattering \cite{NSI-reactor}.

\item 
Solar neutrinos \cite{NSI-solar}.

\item 
Astrophysical neutrinos \cite{NSI-astrophysical}.

\end{itemize}

\section{Acknowledgements}

I deeply thank Milla for her cordial invitation which I hope continues in the future. 
The meeting in Venice epochs my scientific year and 
gives me energy in doing research during the rest of the year. 
I was benefited by useful communications with Enrique Fernandez-Martinez 
and Belen Gavela. 
I thank Christof Wetterich for his comment on my talk in this conference. 
I am grateful to all of my collaborators, in particular, 
Hiroshi Nunokawa, Renata Zukanovich Funchal, 
Shoichi Uchinami, and Takashi Kikuchi for fruitful collaborations. 
This work was supported in part by KAKENHI, Grant-in-Aid for
Scientific Research No. 19340062, 
Japan Society for the Promotion of Science.

\end{document}